\documentclass[10pt,preprint]{aastex}

\newcommand\nion[2]{#1\,\lowercase{{\sc #2}}}
\newcommand\wave[1]{\mbox{$\lambda$#1\,\AA}}
\def\kmsec{${\rm km~s^ {-1}}$}
\def\teff{\mbox{T$_{\rm eff}$}}
\def\BmV0{\mbox{$(B-V)^{\rm 0}$}}
\def\VmK0{\mbox{$(V-K)^{\rm 0}$}}
\def\MV0{\mbox{$M_{\rm V}^{\rm 0}$}}

\def\simgt{\lower.5ex\hbox{$\; \buildrel > \over \sim \;$}}
\def\simlt{\lower.5ex\hbox{$\; \buildrel < \over \sim \;$}}
\def\teff{T_{\rm eff}}
\def\tphot{T_{\rm phot}}
\def\logg{\log~g}
\def\loggb{\log~g_{\rm bol}}
\def\loggp{\log~g_{\rm phot}}

\shorttitle{Origins of the Thick Disk}
\shortauthors{Ruchti et al.}

\begin{document}

\title{Origins of the Thick Disk as Traced by the Alpha-Elements of Metal-Poor Giant Stars Selected from RAVE}

\author{
G.~R.~Ruchti\altaffilmark{1},
J.~P.~Fulbright\altaffilmark{1},
R.~F.~G.~Wyse\altaffilmark{1},
G.~F.~Gilmore\altaffilmark{2},
O.~Bienaym\'{e}\altaffilmark{3}, J.~Binney\altaffilmark{4}, J.~Bland-Hawthorn\altaffilmark{5}, R.~Campbell\altaffilmark{6}, K.~C.~Freeman\altaffilmark{7}, B.~K.~Gibson\altaffilmark{8}, E.~K.~Grebel\altaffilmark{9}, A.~Helmi\altaffilmark{10}, U.~Munari\altaffilmark{11}, J.~F.~Navarro\altaffilmark{12}, Q.~A.~Parker\altaffilmark{13},W.~Reid\altaffilmark{13}, G.~M.~Seabroke\altaffilmark{14}, A.~Siebert\altaffilmark{3}, A.~Siviero\altaffilmark{12,15}, M.~Steinmetz\altaffilmark{15}, F.~G.~Watson\altaffilmark{16}, M.~Williams\altaffilmark{17}, T.~Zwitter\altaffilmark{17,18}
}

\altaffiltext{1}{Johns Hopkins University, 3400 N Charles Street, Baltimore, MD 21218, USA; gruchti@pha.jhu.edu}
\altaffiltext{2}{Institute of Astronomy, University of Cambridge, Madingley Road, Cambridge CB3 0HA, UK}
\altaffiltext{3}{Observatoire de Strasbourg, 11 Rue de L'Universit\'{e}, 67000 Strasbourg, France}
\altaffiltext{4}{Rudolf Pierls Center for Theoretical Physics, University of Oxford, 1 Keble Road, Oxford OX1 3NP, UK}
\altaffiltext{5}{Sydney Institute for Astronomy, School of Physics A28, University of Sydney, NSW 2006, Australia}
\altaffiltext{6}{Western Kentucky University, 1906 College Heights Blvd., Bowling Green, KY 42101, USA}
\altaffiltext{7}{RSAA Australian National University, Mount Stromlo Observatory, Cotter Road, Weston Creek, Canberra, ACT 2611, Australia}
\altaffiltext{8}{Jeremiah Horrocks Institute for Astrophysics \& Super-computing, University of Central Lancashire, Preston,
UK}
\altaffiltext{9}{Astronomisches Rechen-Institut, Zentrum f\"ur Astronomie der Universit\"at Heidelberg, D-69120 Heidelberg, Germany}
\altaffiltext{10}{Kapteyn Astronomical Institute, University of Groningen, Postbus 800, 9700 AV Groningen, Netherlands}
\altaffiltext{11}{INAF Osservatorio Astronomico di Padova, Via dell'Osservatorio 8, Asiago I-36012, Italy}
\altaffiltext{12}{University of Victoria, P.O. Box 3055, Station CSC, Victoria, BC V8W 3P6, Canada}
\altaffiltext{13}{Macquarie University, Sydney, NSW 2109, Australia}
\altaffiltext{14}{Mullard Space Science Laboratory, University College London, Holmbury St. Mary, Dorking, RH5 6NT, UK}
\altaffiltext{15}{Astrophysikalisches Institut Potsdam, An der Sterwarte 16, D-14482 Potsdam, Germany}
\altaffiltext{16}{Anglo-Australian Observatory, P.O. Box 296, Epping, NSW 1710, Australia}
\altaffiltext{17}{Faculty of Mathematics and Physics, University of Ljubljana, Jadranska 19, Ljubljana, Slovenia}
\altaffiltext{18}{Center of Excellence SPACE-SI, Ljubljana, Slovenia}

\begin{abstract}

Theories of thick disk formation can be differentiated by measurements of stellar elemental abundances.  We have undertaken a study of metal-poor stars selected from the RAVE spectroscopic survey of bright stars to establish whether or not there is a significant population of metal-poor thick-disk stars (${\rm [Fe/H]}\simlt-1.0$) and to measure their elemental abundances.  In this paper, we present abundances
of four $\alpha$-elements (Mg, Si, Ca, Ti) and iron for a subsample
of 212 RGB and 31 RC/HB stars from this study.  We find that the [$\alpha$/Fe] ratios are enhanced implying that enrichment proceeded by purely core-collapse supernovae.  This requires that star formation in each star forming region had a short duration.  The relative lack of scatter in the [$\alpha$/Fe] ratios implies good mixing in the ISM prior to star formation.  In addition, the ratios resemble that of the halo, indicating that the halo and thick disk share a similar massive star IMF.  We conclude that the $\alpha$-enhancement of the metal-poor thick disk implies that direct accretion of stars from dwarf galaxies similar to surviving dwarf galaxies today did not play a major role in the formation of the thick disk.

\end{abstract}


\section{Introduction}

Thick disks are a common and significant stellar component in most
disk galaxies, and their formation is an integral part of disk-galaxy
formation.  Most of the stars in the thick disk of the Milky Way
Galaxy are old, $\simgt10$~Gyr \citep{gw85,r06}, and so they can serve
as fossil records of the formation processes in early Galactic
evolution.  An important method for unlocking this information is the
analysis of the elemental abundance patterns in   thick-disk stars.  
Of particular importance are the ratios of the $\alpha$-elemental
(eg. Mg, Si, Ca, Ti) abundances to iron, which provide information
about the past star formation and IMF of a stellar population
\citep[cf.][]{w10}.  This approach is complementary to comparisons of the age
distribution of thick-disk stars with theoretical expectations
\citep[e.g.][]{w09}. 

  Models of
the formation of the thick disk, including scenarios
ranging from migration of stars from the inner disk \citep{sb09} to
heating of the thin disk due to mergers \citep[e.g.][]{villalobos08},
make specific predictions about the chemical abundance properties of
the metal-weak (and oldest) stellar population in the thick disk.  In
hierarchical clustering, heating of pre-existing thin stellar disks by
merging continues until late times.  In this case, an intermediate-age
thick disk would emerge from heating of the thin disk, in conflict with the old
ages found.  A proposed solution is that the old thick disk is a
result of direct accretion of old stars from a few satellite galaxies
\citep{abadi}.  The metal-poor stars in the thick disk would then show
the same chemical enrichment as the parent satellite.
Depending on the mass, orbit, and density profile of each accreted
satellite, the $\alpha$-abundance patterns of the thick disk would
also vary with Galactic radius.  The identification and chemical
analysis of metal-poor thick-disk stars beyond the solar neighborhood
is therefore very important.

The existence of a significant metal-weak (${\rm [Fe/H]}\leq-1.0$)
thick disk remains controversial.  The metal-poor thick disk was
identified in both the field \citep{n85, m90,wg95, cb00}, and in
globular clusters \citep{d99}. The apparent lack of blue
horizontal-branch and RR Lyrae stars in the thick disk, however,
argues against an old metal-poor component \citep{k09}.  The small
number of metal-poor thick-disk stars studied to date \citep{f02,b03,
bc06, r06,reddy08} have high [$\alpha$/Fe] ratios, implying they
formed in a short-duration star formation event, and old ages when
estimates are available.  These samples, however, have been limited to
very bright stars within the solar neighborhood.

The unprecedented size of the Radial Velocity Experiment spectroscopic
survey \citep[RAVE,][]{rave} and its selection without either
kinematic or metallicity criteria provide a unique opportunity to
study a statistically significant sample of bright metal-poor
thick-disk stars, as described below.  Our sample both probes a much
larger volume and extends to lower metallicities than any previous
thick-disk sample with elemental abundances.  We here present
elemental abundance results for a subsample of (predominantly) red
giant branch (RGB) and red clump/horizontal branch (RC/HB) stars.

\section{Observations}
\subsection{Candidate Selection}
\label{sec-cand}

Candidate metal-poor thick-disk stars were selected from RAVE, a {\it
magnitude-limited} survey that uses the 6dF spectrograph on the UK
Schmidt telescope to obtain high S/N, $\mathcal{R}\sim7500$ spectra
(\wave{8410}--\wave{8795}) of stars in the southern sky with $I<13$.
The RAVE pipeline provides estimates of radial velocities and stellar
parameters (effective temperature, $\teff$; gravity, $\logg$;
metallicity, [M/H]) through fits to a grid of synthetic spectra
\citep[see][]{z08}.  The stars are bright enough that they have proper
motion measurements in the literature (included in the RAVE database).
We selected candidate metal-poor disk stars with ${\rm [M/H]}\leq-0.7$
and estimates of 3-D space motions, based on parameter values in the
database, consistent with disk kinematics.

\subsection{High Resolution Echelle Observations}

High resolution spectroscopy provides robust abundances for most
elements.  We obtained data between May 2007 and February 2009 using
the following spectrographs/telescopes: MIKE/Magellan-Clay,
FEROS/MPG 2.2-m, UCLES/AAT, and ARCES/APO-3.5.  All instruments have a
resolving power between 35,000-45,000 and a spectral coverage of
$\wave{3500}-\wave{9500}$, except for UCLES which covers
$\wave{4460}-\wave{7270}$.  The raw FEROS spectra were reduced using
the Data Reduction System within ESO-MIDAS, while all other spectra
were reduced using the echelle package in IRAF\footnote{Distributed by NOAO, operated by AURA under cooperative agreement with
the NSF.}. The final spectra yielded a ${\rm S/N}>100$ per pixel at
$\wave{5000}-\wave{6000}$ and a minimum ${\rm S/N}\sim40$ around
\wave{4000}, sufficient for abundance analysis.
    
A total of $\sim500$ spectra for candidate metal-poor thick-disk stars
were obtained.  We removed potentially problematic stars (e.g. hot
stars, fast rotators), in addition to 4 radial-velocity outliers, probable binary
stars, after comparison with RAVE ($\Delta
V>20$~\kmsec).  Ten of the remaining stars have repeat echelle
observations.  We here  report on the subsample of evolved
stars ($\log~g<3.3$), consisting of 212 RGB stars and 31 RC/HB stars,
for which a uniform analysis is implemented.

\section{Abundance Analysis}
\label{abs}

Our initial analysis followed the methodology of \citet{f00}.  We
combined the line lists from F00 and \citet{jj02}, thus 
extending the analysis to extremely metal-poor stars.  Equivalent
widths (EWs) were measured using the ARES code \citep{ares}, which
fits a Gaussian to each line.  Strong lines with ${\rm
EW}\geq110$~m\AA~were removed to minimize poor fits.  Comparisons with
hand-measured EWs for 12 metal-poor stars resulted in a mean
difference ${\rm EW_{hand}-EW_{ARES}}\sim-0.1\pm3$~m\AA, sufficiently small 
that line-measurement biases are unimportant.

The MOOG analysis program \citep{moog} was utilized in an iterative
procedure to compute elemental abundances using 1-D, LTE,
plane-parallel Kurucz model atmospheres,\footnote{http://kurucz.harvard.edu/.}.  The stellar temperature was set by the
excitation temperature method based on \nion{Fe}{I} lines.  The
$\logg$ value was set by minimizing the difference between the
calculated abundance of iron from the \nion{Fe}{I} and \nion{Fe}{II}
lines.  A microturbulent velocity was selected to minimize the slope
of the relationship between the iron abundance derived from
\nion{Fe}{I} lines and the value of the reduced width of the line.
The metallicity value of the stellar atmosphere was chosen to match
the iron abundance in the analysis.  
The repeat observations gave internal errors of 55~K, 0.1~dex, and
0.06~dex in $\teff$, $\logg$, and [Fe/H], respectively, consistent
with those of F00.

External errors are critical in distance determination.  We tested
the accuracy of our parameters using echelle data (obtained for
another project) of several globular cluster RGB stars ($\log~g<2$)
for which the distances and metallicities are reported in the 
literature.  We also reanalyzed a sample (hereafter F00/Hip) of 6
giant stars from F00 which had Hipparcos parallaxes with acceptable
errors \citep[$\sigma_{p}/p<0.2$;][]{vanleeuwen07}.  We found that
with our initial
parameter values many globular cluster giants lay above the RGB-tip, 
contrary to their location in the CMD, implying that
our initial  $\logg$ estimate is too low.
Further, the interplay between $\logg$ and $\teff$ in our initial
analysis implies  that our $\teff$ estimate would also be suspect.  We
therefore computed independent estimates of both gravity and
temperature to quantify these effects and hence correct them.

The independent estimate of surface gravity for each globular cluster and F00/Hip star was obstained using the equation $g_{\rm bol}=4\pi GM\sigma \teff^4/L$.  We adopt our initial estimate of $\teff$, a typical RGB star mass of $0.8~{\rm M_\odot}$, and estimated the luminosity from the de-reddened 2MASS $K_{\rm S}$ with bolometric corrections derived from \citet{ghernandez09} and the published distance.  We also derived an independent photometric temperature, $\tphot$, for each star using the 2MASS color-temperature transformations from \citet{ghernandez09}.  We assumed that the scale provided by $\tphot$ does not show spurious trends such as apparently introduced by our initial spectroscopic analysis, and thus can be used to correct our $\logg$ estimates (as achieved below).  

Comparisons between our spectroscopic estimates and these derived values revealed offsets in both temperature and gravity.  Further, the difference between our spectroscopic $\teff$ estimates and $\tphot$ is correlated with iron abundance such that our estimate is $300-400$~K cooler for the lowest metallicity stars (note both \citealt{jj02} and \citealt{a05} found similar offsets, albeit using different photometric colors in the derivation of $\tphot$).  We therefore corrected $\teff$ according to this correlation and repeated the analysis to obtain a new (ionization-balanced) gravity estimate, $\loggp$.  Stars with $\loggp\geq1$ showed no mean offset with $\loggb$ and we therefore adopted $\loggp$ as our final gravity.  The difference, $\loggp-\loggb$ is however correlated with $\loggp$ for $\loggp<1$.  For these stars, we adopt a new $\logg$ estimate from this correlation along with the corrected temperature above to get a final estimate of the iron abundance.  The final temperatures and gravities showed scatter with respect to $\tphot$ and $\loggb$ of 140~K and 0.2~dex, respectively.  In the cases when the iron abundance from \nion{Fe}{I} and \nion{Fe}{II} do not agree, we chose \nion{Fe}{II} as our final estimate, since \nion{Fe}{II} is both the dominant species and much less sensitive to non-LTE effects than \nion{Fe}{I} \citep{ti99,as99}.  The iron abundances derived for any given globular cluster showed a star-to-star scatter of $\pm0.1$~dex, consistent with our earlier estimates of our internal errors.  

We used the above procedure to insure that the derived parameter values for our RAVE sample provide accurate distances.

\section{Sample Analysis}
\label{sec-an}
\subsection{Distances}

Distances are critical and a full description of our technique is given in Ruchti et al. (2010, in prep).  In short, distances to our stars are estimated from the most probable $M_{K_{\rm S}}$ magnitude in a grid of Padova isochrones \citep{m08,piso}.   The temperature, gravity, and metallicity, following the procedure in \S3, and their associated errors (as described above) of 140 K, 0.2 dex, and 0.1 dex in $\teff$, $\logg$, and [Fe/H], respectively, are used in the multi-parameter fit.  For the RGB stars, our technique weights each point by the luminosity function, derived from the BaSTI luminosity function tracks \citep[these isolate the RGB from the AGB;][]{p04}, to emphasize evolutionary stages with longer lifetimes.  Several of our stars have $\teff$ and $\logg$ inconsistent with the RGB, and appear to be RC/HB stars.  The position of a star on the horizontal branch depends on the mass-loss on the RGB, which is not well understood or modeled.  The Padova isochrones, which we use, represent the zero-age HB as one point.  Stars within $2\sigma_{\teff}$ of this point were assumed to have an absolute magnitude equal to that of that point.  Note that $M_{Ks}$ ranges between only $0.1-0.2$ magnitudes during the slow-evolution phase of core-helium burning.  We assumed an old age for the RGB stars, performing a weighted average over isochrones of ages 10, 11, and 12~Gyr (reducing the age to 5~Gyr increased the distances by only 10\%), while we assume an age of 12~Gyr for the RC/HB stars. 

We first applied this technique to the cluster and F00/Hip stars.  Our
distances differed from literature values by only $-2\pm15\%$ for the
cluster stars and $-4\pm13\%$ for the F00/Hip stars.  We therefore
adopted a conservative estimate of 20\% error on the distance,
including both scatter and offset.  Distance estimates based on RAVE
pipeline values of stellar parameters are now
available\citep{breddels10,z10}.  Our distance estimates for the 172
stars with $|\logg_{\rm echelle}-\logg_{\rm RAVE}|<0.5$ are shorter
than those of Zwitter et al. by $15\pm24\%$.  Note that our technique
was optimized for metal-poor stars with echelle parameters, while
\citet{z10} optimized their method for all stars in the RAVE catalog,
which have a high mean metallicity and typically younger ages.

The average distance to our RAVE stars is $\sim2$~kpc; all within
$\sim7$~kpc, except one at $\sim16$~kpc.  The majority of our stars
have Galactic longitudes between $\ell=200^{\circ}$ and $50^{\circ}$
and Galactic latitudes $|b|>25^{\circ}$, extending to an average
vertical height of $|z|\simgt1$~kpc.

\subsection {Population Assignments}

The 3-D, cylindrical space motion ($V_{\Pi}$, $V_\Theta$, $V_Z$) of
each star was computed by combining the derived distances and radial
velocities with proper motions from the RAVE database.  We adopted (10, 5.25, 7.17)~\kmsec \citep{db98} for the solar
motion\footnote{Our assignment method is insensitive to
values within the range of recent estimates
\citep[e.g.][]{sch10}.}  and $V_{\rm
LSR}=220$~\kmsec. Each component of a star's space motion was
then re-sampled 10,000 times, assuming a normal distribution centered
on our estimate of the component velocity, with standard deviation
equal to the propagated error in the velocity.  We computed the
probability that each re-sampled value of the space motion was drawn from a given kinematic population, based on the combined local characteristic
Gaussian distributions assumed for each Galactic population (see Table 
\ref{tab-gaus}), defined as:
\begin{equation}
P(V_{\Pi}, V_{\Theta}, V_{Z}) \propto \exp \left( -\frac{V_{\Pi}^2}{2\sigma_{\Pi}^2} - \frac{(V_{\Theta} - 220 - \langle V_{\Theta} \rangle)^2}{2\sigma_{\Theta}^2} - \frac{V_Z^2}{2\sigma_{Z}^2} \right),
\label{eq-gaus}
\end{equation}
where, for each population, $\sigma_{\Pi}$, $\sigma_{\Theta}$, and $\sigma_{Z}$ are the
characteristic velocity dispersions and $\langle V_{\Theta} \rangle$
is the mean rotational velocity (corrected for the Sun's motion).  Note that Gaussian distributions are a first-order approximation, and the distribution functions may be much more complex \citep[cf.,][]{b10}.   We did not include vertical gradients in kinematics or metallicity-kinematical relations at this point.

\begin{table}[h]
\caption{Local Characteristic Velocity Distributions.}
\begin{tabular}{lrrrrl}
\hline
\hline& $\sigma_{\Pi}$ & $\sigma_{\Theta}$ & $\sigma_{Z}$ & $\langle V_{
\Theta} \rangle$ & Ref.\\
& (\kmsec) & (\kmsec) & (\kmsec) & (\kmsec) & \\
\hline
Thin Disk & 39 & 20 & 20 & -15 & \citet{soub} \\
Thick Disk & 63 & 39 & 39 & -51 & \citet{soub}\\
Halo  & 141 & 106 & 94 & -220 & \citet{cb00} \\
\hline
\end{tabular}
\label{tab-gaus}
\end{table}

For each of the 10,000 samplings of the space motion of a given star, an assignment to the thick disk was made if that 
probability was 4 times that of either the thin disk or halo, with an
analogous procedure for thin disk and halo.  Intermediate populations,
thin/thick and thick/halo, were assigned for lower values of
probability ratios.  A star was assigned to the Galactic population
with the highest number of occurrences from the 10,000 re-sampled
points.

A second population assignment was determined by comparing a star's
position in the Galaxy to the characteristic density distributions of
each Galactic component \citep[from][]{juric08}, following the same
Monte Carlo procedure as the space motion criterion.  This criterion
works well far from the Galactic plane, however discrimination is
difficult close to the plane since a star must have $|z|=2$~kpc before
the thick disk probability is 4 times that of the thin disk.  The
assignment is therefore restricted to choosing only halo over thick
disk or thick over thin disk.  

\begin{figure}[h]
\epsscale{1.0}
\plotone{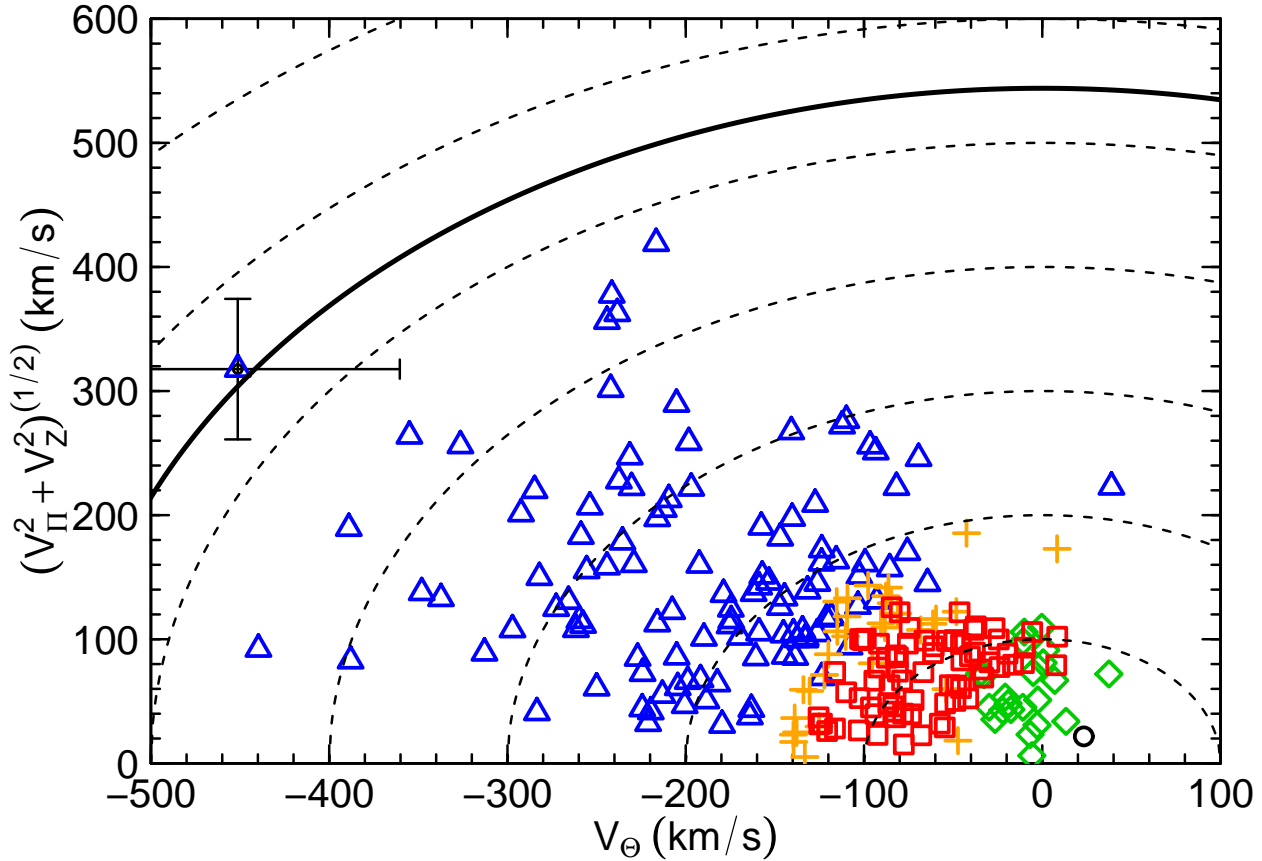}
\caption{The Toomre Diagram for our sample with
$\sigma_{V_{\theta}}<100$~\kmsec.  The black circle, green diamonds,
red squares, orange plus signs, and blue triangles correspond to thin
disk, thin/thick, thick disk, thick/halo, and halo stars,
respectively.  The dashed curves indicate constant space motion, in
steps of 100~\kmsec.  The thick solid curve is an estimate of the local 
escape velocity \citep{s07}.  The total
velocity error depends on distance and proper motion errors, with the typical $1\sigma$ error  $<20$~\kmsec.  The star near the estimated escape velocity is the subject of a future paper.}
\label{fig-toom}
\end{figure}

The final population assignment was then the result from the space
motion criterion, unless overridden by the boundary condition set from
the star's position.  Our results are consistent with each star's
position in the reduced proper motion diagram.  These final
assignments were 73 thick disk, 1 thin disk, 22 thick/thin, 31
thick/halo, and 116 halo.  The Toomre diagram (Figure~\ref{fig-toom}) 
illustrates the correspondence between these population assignments and velocities.  It is not
surprising that many candidates were assigned to the halo since, with
our final parameter values, our sample extends to large distances and
high velocities.

\section{Alpha Element Abundance Results}
\label{sec-alphas}

\begin{figure}[h]
\epsscale{0.8}
\plotone{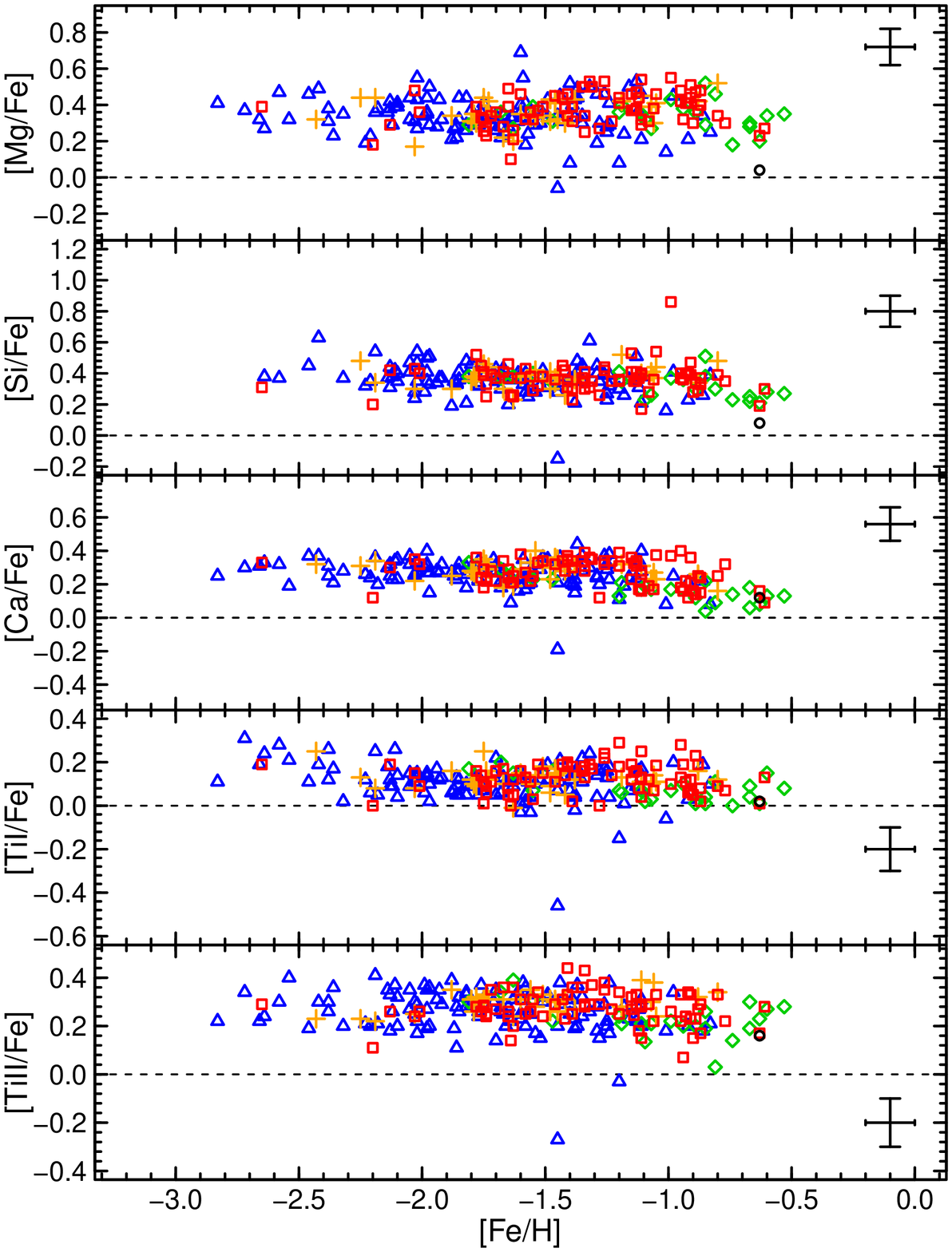}
\caption{Computed [$\alpha$/Fe] ratios versus [Fe/H] for our sample of stars.  While \nion{Fe}{II} is used to estimate [Fe/H], element ratios are computed using the iron abundance of the same ionization state as the $\alpha$-element (e.g. [Si/Fe] = [\nion{Si}{I}/\nion{Fe}{I}]) as is suggested by \citet{ki03}.  Color and symbols are the same as in Fig.~\ref{fig-toom}.  The cause of the offset between [\nion{Ti}{I}/Fe] and [\nion{Ti}{II}/Fe] is unclear, but the thick disk and halo still show similar enhancement in each.  Two stars of note are the halo star at ${\rm [Fe/H]}\sim-1.45$ with very low $\alpha$-enhancement and the thick disk star at ${\rm [Fe/H]}\sim-1$ with a very high [Si/Fe] ratio, each of which is the subject of future papers.}
\label{fig-alf}
\end{figure}

Elemental abundances were derived through MOOG after all stellar parameter corrections in \S3.  Figure~\ref{fig-alf} displays [$\alpha$/Fe] as a function of [Fe/H] for several $\alpha$-elements.  Note that the range of [Fe/H] shows that our selection function from RAVE was efficient.  The main bulk of thick-disk stars extend to ${\rm [Fe/H]}\sim-1.8$, with 5 additional stars assigned to the thick disk below -2~dex.  Further, these stars have  [$\alpha$/Fe] significantly above solar, with relatively low scatter, and blend smoothly into the halo stars. 

The kinematics of the metal-poor thick disk may differ from the canonical thick disk \citep{c09}, with higher velocity dispersions and a slower rotational velocity.  Adopting $\langle V_{\Theta} \rangle=-100$~\kmsec~\citep[see][]{g02}, would reassign several stars with thick/halo population assignments to the thick disk.  There would be no change in our conclusions, since these stars have similar $\alpha$-enhancement to the metal-poor thick disk and halo.

A two-component halo at ${\rm [Fe/H]}\simgt-1$, separated in [Mg/Fe] enhancement, has been reported by \citet{ns10}.  Indeed, several of our halo stars in this metallicity range have values of ${\rm [Mg/Fe]}\simlt0.2$, but other $\alpha$-elements show little spread, in contrast to \citet{ns10}.  We will return to this in another paper.

\section{Discussion}

We have computed the abundances of four major $\alpha$-elements for RGB and RC/HB stars in our sample, with $-0.5\simgt{\rm[Fe/H]}\simgt-2.8$.  About 40\% of these stars have the highest probability of being (thick or thin) disk stars and probe distances much further than any earlier investigation.  Previous high-resolution samples \citep[e.g.][]{reddy08} did identify metal-poor thick disk stars with ${\rm[Fe/H]}<-1.5$.  We have increased this number by almost an order of magnitude, finding 24 thick disk stars with ${\rm[Fe/H]}<-1.5$ that extend to metallicities below -2~dex.   The [$\alpha$/Fe] ratios for these stars are enhanced, indicating that the metal-poor thick disk was enriched by SNe II, before SNe Ia enrichment, implying a timescale for star formation shorter than $\sim1$~Gyr.  The relatively low scatter in [$\alpha$/Fe] ratios also implies that the ISM was well mixed prior to star formation, and the metal-poor thick-disk stars were enriched by supernovae from an invariant massive star IMF.  Further, the metal-poor thick disk and halo stars show similar $\alpha$-enhancement, evidence that they were pre-enriched by the same massive star IMF.    

The $\alpha$-enhancement in the metal-poor thick disk contrasts with that seen in stars of similar metallicities in local dwarf galaxies \citep[as reviewed by][]{t09}.  All local dwarf galaxies began star formation early, $10-12$~Gyr ago, and had extended star formation, often with dominant intermediate-age populations \citep[e.g.][]{t09}.  Direct accretion of stars by assimilation of dwarf satellite galaxies into the thin and thick disks in the \citet{abadi} models lasts until $z\sim0.7$ ($\sim6$~Gyr ago).  If the accreted dwarf galaxies formed stars until accretion or had extended star formation similar to surviving dwarfs, then many of the accreted stars will have formed from gas which had significant iron contribution from SNe Ia.  Consequently, stars accreted from dwarfs into the thick disk after this time will have low [$\alpha$/Fe] ratios \citep{u96}, contrary to the enhancement we see in our sample.  We conclude that direct accretion of stars from dwarf galaxies similar to surviving dwarf galaxies today did not play a major role in the formation of the thick disk.

Another possibility is that the thick disk was formed from multiple gas-rich minor mergers, as simulated by \citet{brook}.  This model predicts a high star formation rate from the gas dissipation, consistent with our enhanced [$\alpha$/Fe], but the model includes direct accretion of stars from the satellite galaxies.  We would therefore still expect to find some low [$\alpha$/Fe] stars in the thick disk from {\it late} merging, which conflicts with our low scatter in the abundance ratios of the metal-poor thick disk.  Early heating of a thin stellar disk by mergers, however, is still viable.

The amplitude of radial and vertical gradients in abundances and metallicity will provide diagnostics to further discriminate models of the formation of the thick disk.  Our sample, which probes distances much further from the solar neighborhood, is the first for which this can be done, and will be investigated in our next paper (Ruchti et al. 2010, in prep).  

\acknowledgements
GRR, JPF, and RFW acknowledge support through grants from the W. M. Keck Foundation and the Gordon and Betty Moore Foundation, to establish a program of data-intensive science at the Johns Hopkins University, as well as from the NSF (AST-0908326).  This publication makes use of data products of the 2MASS survey, a joint project of the University of Massachusetts and IPAC/Caltech, funded by NASA and the NSF. Funding for RAVE (www.rave-survey.org) has been provided by institutions of the RAVE participants and by their national funding agencies.

{\it
\facility
Facilities: \facility{ARC (echelle spectrograph)}, \facility{AAT (UCLES)}, \facility{Magellan:Clay (MIKE)}, \facility{Max (FEROS)}, \facility{UKST (6dF spectrograph)}
}

\clearpage


\begin{thebibliography}{}
\bibitem[Abadi et al.(2003)]{abadi} Abadi, M. G., Navarro, J. F., Steinmetz, M. \& Eke, V. R. 2003, \apj, 597, 21
\bibitem[Aoki et al.(2005)]{a05} Aoki, W., et al.\ 2005, \apj, 632, 611 
\bibitem[Asplund et al.(1999)]{as99} Asplund, M., Nordlund, {\AA}., Trampedach, R., \& Stein, R.~F.\ 1999, \aap, 346, L17
\bibitem[Bensby et al.(2003)]{b03} Bensby, T., Feltzing, S. \& Lundstrom, I. 2003, A\&A, 410, 527
\bibitem[Binney(2010)]{b10} Binney, J.\ 2010, \mnras, 401, 2318
\bibitem[Breddels et al.(2010)]{breddels10} Breddels, M.~A., et al.\ 2010, \aap, 511, A90
\bibitem[Brewer \& Carney(2006)]{bc06} Brewer, M.-M. \& Carney, B. W. 2006, \aj, 131, 431
\bibitem[Brook et al.(2005)]{brook} Brook, C. B., Gibson, B. K., Martel, H. \& Kawata, D. 2005, \apj, 630, 298
\bibitem[Dehnen \& Binney(1998)]{db98} Dehnen, W., \& Binney, J.~J.\ 1998, \mnras, 298, 387
\bibitem[Dinescu et al.(1999)]{d99} Dinescu, D. I., Girard, T. M. \& van Altena, W. F. 1999, \aj, 117, 1792
\bibitem[Carollo et al.(2010)]{c09} Carollo, D., et al. 2010, \apj, 712, 692
\bibitem[Chiba \& Beers(2000)]{cb00} Chiba, M. \& Beers, T. C. 2000, \aj, 119, 2843
\bibitem[Fulbright(2000, hereafter F00)]{f00} Fulbright, J. P. 2000, \aj, 120, 1841 (F00)
\bibitem[Fulbright(2002)]{f02} Fulbright, J. P. 2002, \aj, 123, 404
\bibitem[Genzel et al.(2006)]{genzel} Genzel, R. et al. 2006, Nature, 442, 786
\bibitem[Gilmore \& Wyse(1985)]{gw85} Gilmore, G., \& Wyse, R.~F.~G.\ 1985, \aj, 90, 2015
\bibitem[Gilmore et al.(2002)]{g02} Gilmore, G., Wyse, R. F. G. \& Norris, J. E. 2002, \apj, 574, 39
\bibitem[Girardi et al.(2002)]{piso} Girardi, L. et al. 2002, A\&A, 391, 195
\bibitem[Gonz{\'a}lez Hern{\'a}ndez \& Bonifacio(2009)]{ghernandez09} Gonz{\'a}lez Hern{\'a}ndez, J.~I., \& Bonifacio, P.\ 2009, \aap, 497, 497
\bibitem[Johnson(2002)]{jj02} Johnson, J. A. 2002, ApJS, 139, 219
\bibitem[Juri{\'c} et al.(2008)]{juric08} Juri{\'c}, M., et al.\ 2008, \apj, 673, 864
\bibitem[Kinman et al.(2009)]{k09} Kinman, T. D., Morrison, H. L., \& Brown, W. R., 2009, \aj, 137, 3198
\bibitem[Kraft \& Ivans(2003)]{ki03} Kraft, R.~P., \& Ivans, I.~I.\ 2003, \pasp, 115, 143
\bibitem[Marigo et al.(2008)]{m08} Marigo, P. et al. 2008, A\&A, 482, 883
\bibitem[Morrison et al.(1990)]{m90} Morrison, H. L., Flynn, C. \& Freeman, K. C. 1990, \aj, 100, 1191
\bibitem[Nissen \& Schuster(2010)]{ns10} Nissen, P.~E., \& Schuster, W.~J.\ 2010, \aap, 511, L10
\bibitem[Norris et al.(1985)]{n85} Norris, J., Bessell, M. S. \& Pickles, A. J. 1985, ApJS, 58, 463 
\bibitem[Pietrinferni et al.(2004)]{p04} Pietrinferni, A., Cassisi, S., Salaris, M., \& Castelli, F.\ 2004, \apj, 612, 168
\bibitem[Reddy et al.(2006)]{r06} Reddy, B. E., Lambert, D. L. \& Allende Prieto, C. 2006, MNRAS, 367, 1329
\bibitem[Reddy \& Lambert(2008)]{reddy08} Reddy, B.~E., \& Lambert, D.~L.\ 2008, \mnras, 391, 95
\bibitem[Sch{\"o}nrich \& Binney(2009)]{sb09} Sch{\"o}nrich, R., \& Binney, J.\ 2009, \mnras, 396, 203
\bibitem[Sch{\"o}nrich et al.(2010)]{sch10} Sch{\"o}nrich, R., Binney, J., \& Dehnen, W.\ 2010, \mnras, 403, 1829
\bibitem[Smith et al.(2007)]{s07} Smith, M. et al. 2007, MNRAS, 379, 755
\bibitem[Sneden(1973)]{moog} Sneden, C. 1973, \apj, 184, 839
\bibitem[Soubiran et al.(2003)]{soub} Soubiran, C., Bienaym\'{e}, O. \& Siebert, A. 2003, \aap, 398, 141
\bibitem[Sousa et al.(2007)]{ares} Sousa, S. G., Santos, N. C., Israelian, G., Mayor, M. \& Monteiro, M. J. P. F. G. 2007, A\&A, 469, 783
\bibitem[Steinmetz et al.(2006)]{rave} Steinmetz, M.  et al. 2006, \aj, 132, 1645
\bibitem[Th{\'e}venin \& Idiart(1999)]{ti99} Th{\'e}venin, F., \& Idiart, T.~P.\ 1999, \apj, 521, 753
\bibitem[Tolstoy et al.(2009)]{t09} Tolstoy, E., Hill, V., Tosi, M. 2009, ARA\&A, 47, 371
\bibitem[Unavane et al.(1996)]{u96} Unavane, M., Wyse, R.~F.~G., \& Gilmore, G.\ 1996, \mnras, 278, 727
\bibitem[van Leeuwen(2007)]{vanleeuwen07} van Leeuwen, F.\ 2007, \aap, 474, 653
\bibitem[Villalobos \& Helmi(2008)]{villalobos08} Villalobos, {\'A}., \& Helmi, A.\ 2008, \mnras, 391, 1806
\bibitem[Wyse \& Gilmore(1995)]{wg95} Wyse, R.~F.~G., \& Gilmore, G.\ 1995, \aj, 110, 2771
\bibitem[Wyse(2009)]{w09} Wyse, R.~F.~G.\ 2009, IAU Symposium, 254, 501
\bibitem[Wyse(2010)]{w10} Wyse, R.~F.~G.\ 2010, IAU Symposium, 265, 461
\bibitem[Zwitter et al.(2008)]{z08} Zwitter, T., et al. 2008, \aj, 136, 421
\bibitem[Zwitter et al.(2010)]{z10} Zwitter, T., et al. 2010, A\&A, accepted, arXiv:1007.4411 
\end{thebibliography}
\end{document}